\documentclass[manuscript]{acmart}
\AtBeginDocument{%
  }

\setcopyright{none}
\acmDOI{}
\acmISBN{}

\acmConference[CHI '26 Workshop]{CHIdeology}{April 15, 2026}{Barcelona, Spain}

\begin{document}

\title{Building a ``-Sensitive Design'' Methodology from Political Philosophies or Ideologies}


\author{Anthony Maocheia-Ricci}
\affiliation{%
 \institution{University of Waterloo}
 \city{Waterloo}
 \state{Ontario}
 \country{Canada}}

\author{Edith Law}
\affiliation{%
 \institution{University of Waterloo}
 \city{Waterloo}
 \state{Ontario}
 \country{Canada}}

\renewcommand{\shortauthors}{Maocheia-Ricci and Law}

\begin{abstract}
Value-based approaches such as Value Sensitive Design (VSD) enable technology designers to engage with and integrate human values in technology through a tripartite methodology of conceptual, empirical, and technical investigations. However, VSD contains pitfalls in both translating values to requirements and a lack of normative grounding, leading to adaptations such as Jacobs' Capability Sensitive Design (CSD). Inspired by CSD and extensions of the design approach, we propose the concept of creating -Sensitive Design (-SD); a meta-framework to embed various political or ideological values as norms in a design research process. We exemplify this through \emph{Dependency}-Sensitive Design (DSD), combining ideas from Kittay's critiques of classical liberal theory within a practical VSD framework. Finally, we push for further work combining philosophy and design in areas beyond CSD and DSD.
\end{abstract}



\maketitle

\section{Value- and Capability-Sensitive Design}
Much work in recent decades has gone towards crafting design methodologies with the aim to integrate human values in technology; namely Friedman and Hendry~\cite{friedmanValueSensitiveDesign2019}, Friedman et al.~\cite{friedmanValueSensitiveDesign2013}, Flanagan et al.~\cite{article}, van de Hoven~\cite{vandenhovenICTValueSensitive2007}, and van de Poel~\cite{vandepoelTranslatingValuesDesign2013, vandepoelDesignValueChange2021}. Of note is \emph{Value Sensitive Design} (VSD) and its adaptations, a framework for designing technology where the process is grounded in human values of interest for a particular group of stakeholders. VSD is described by Friedman et al.~\cite{friedmanValueSensitiveDesign2013} as an iterative \emph{tripartite} methodology in which researchers conduct (1) \emph{conceptual} investigations on both values and relevant stakeholders from prior literature, (2) \emph{empirical} investigations through any range of qualitative or quantitative methodologies used in human studies, and (3) \emph{technical} investigations of existing or novel technology created through the VSD process.


However, scholars have pointed out some pitfalls found within VSD. First, VSD does not include a concrete procedure to translate values into \emph{design requirements}. To resolve this, van de Poel~\cite{vandepoelTranslatingValuesDesign2013} introduces the \emph{value hierarchy}, where high-level values are first translated into prescriptive \emph{norms}, related to objectives, goals, or constraints a technology should have, prior to translation into specific design requirements. Alternatively, the hierarchy may be constructed in a bottom-up fashion, going from concrete requirements to abstract values. Beyond procedural problems, Manders-Huits~\cite{manders-huitsWhatValuesDesign2011} describes the openness of `value' within standard approaches to VSD as lacking a normative justification. As a result of lacking clear empirical standards and a relativist understanding of value definitions within study populations, Manders-Huits argues for the inclusion of an ethical framework or theory when practising VSD to provide a baseline understanding of morals separate from stakeholders and designers. To actualize this, Jacobs~\cite{jacobsCapabilitySensitiveDesign2020} augmented Friedman and Hendry's conception of VSD~\cite{friedmanValueSensitiveDesign2019} with Nussbaum's Capability Theory~\cite{nussbaumCapabilitiesFundamentalEntitlements2006} to create \emph{Capability Sensitive Design} (CSD).

Nussbaum's Capability Theory, modelled after Sen's Capability Approach (CA)~\cite{senInequalityReexamined1992a}, is a feminist theory of social justice discussing fundamental rights every human \emph{ought} to have to live a dignified life~\cite{nussbaumCapabilitiesFundamentalEntitlements2006}. Nussbaum constructs an open-ended list of 10 central capabilities, including concepts of \emph{Life}, \emph{Bodily Health}, \emph{Emotions}, and others. Jacobs, acknowledging VSD's pitfall in normativity, treats Nussbaum's 10 capabilities as a normative framework for values in the context of social justice, health, and well-being technologies. Following the same iterative triparite methodology of VSD, CSD involves engaging with these capabilities in conceptual, empirical, and technological investigations. Going beyond standard VSD, Jacobs includes van de Poel's value hierarchy~\cite{vandepoelTranslatingValuesDesign2013} in line with Oosterlaken~\cite{oosterlakenHumanCapabilitiesDesign2015} by default in the process, arguing for the use of a \emph{capability hierarchy} in technical investigations with CSD. 

As of yet, minimal work has actualized Jacobs' CSD in research. van der Veen et al.~\cite{vanderveenDesigningBiobasedValue2024} adapted CSD in designing bio-based value chains, where agricultural residues are transformed into alternative products (e.g., biofuels), within the context of social justice. Operationalizing CSD further, Maocheia-Ricci et al.~\cite{maocheia-ricciAccountingDisadvantagesCapability2026} both introduce missing concepts of \emph{corrosive disadvantages} and \emph{fertile functionings} from adaptations of capability theory to CSD~\cite{wolffDisadvantage2007} to tackle relations between capabilities and introduce a standard workflow, \emph{CSD-Cascade}, to use CSD in practice.

This previous work in values-based approaches, VSD, CSD, and CSD-Cascade point towards the power of grounding values-based approaches in the norms of various political philosophies for work in or with particular contexts or communities. In line with Jacobs' work in adapting capabilties to create CSD, we introduce a general meta-framework for creating a -Sensitive Design, or \emph{-SD}.

\section{Creating a -Sensitive Design: \emph{Dependency}-Sensitive Design}
To motivate the creation of a -SD, we first describe in brief Eva Kittay's \emph{dependency} critique of Rawls' liberal theory and ethics of care to use as a normative basis in a theoretical -SD. Rawls' Theory of Justice~\cite{rawlsTheoryJusticeRevised1999} contains five primary goods that act as a basis for equality in society, namely:
{\it\begin{enumerate}
    \item Freedom of thought and conscience,
    \item Freedom of movement,
    \item Powers or offices granting a degree of governance to individuals,
    \item Income and wealth,
    \item and Social bases of self-respect~\cite{rawlsKantianConstructivismMoral1980}.
\end{enumerate}}
In \emph{Love's Labour}~\cite{kittay1999love}, Kittay argues that Rawls' list of primary goods is incomplete due to it not adequately accounting for the needs of dependants and care workers. As such, she proposes a sixth primary good of \emph{Being cared for as a dependant and supported when we care for dependants}. Focusing on dependency and care, Kittay's ethics focuses primarily on justice with respect to people with disabilities or in a period of dependency (e.g., older adults)~\cite{kittayEthicsCareDependence2011}.

As with both Friedman and Hendry's VSD~\cite{friedmanValueSensitiveDesign2019} and Jacobs' CSD~\cite{jacobsCapabilitySensitiveDesign2020}, a -SD approach would follow the same tripartite methodology of investigations. Adapting this Kittay's work, we create \emph{Dependency}-Sensitive Design (DSD) under the same tripartite workflow, CSD-Cascade, outlined in Maocheia-Ricci et al.~\cite{maocheia-ricciAccountingDisadvantagesCapability2026}. In CSD-Cascade, the conceptual investigation begins with community identification and the rewording of capabilities with respect to the community. For a novel -SD, a philosophical or ideological framework should be identified in tandem; a hypothetical DSD would better fit a technology design process for older adults or those with physical disabilities over newcomers or refugees. As such, a conceptual investigation for creating an -SD approach should be augmented with an \emph{ideological} investigation to discover a set of concrete values, rooted in ideology or philosophy, matching the stakeholders or community of focus. After both are discovered, the values should be initially reworded with respect to the particular community at hand; in DSD, the primary goods could be simplified to: (1) \emph{Agency in thought}, (2) \emph{Agency in movement and travel}, (3) \emph{Agency in self-governance}, (4) \emph{Ability to meet ends}, (5) \emph{Respect from social institutions}, and (6) \emph{Care for dependants and caregivers}.

After an initial conceptual investigation, empirical investigations should take place with both direct (community) and indirect stakeholders. These investigations should make use of the values outlined in a -SD (e.g., primary goods for DSD) in ways such as Steen~\cite{steenOrganizingDesignforWellbeingProjects2016} or Javornik et al.'s~\cite{javornikAskRatherAssume2019a} capability cards for workshops (as used in interviews and co-design by Maocheia-Ricci et al.~\cite{maocheia-ricciAccountingDisadvantagesCapability2026}). From these investigations, various factors should be uncovered, including priorities and gaps within values, functions used to realize values in practice, and the interconnectedness of values themselves. CSD realizes the former two through the concept of \emph{conversion factors}, or what turns a capability as goal into \emph{functioning} as ability. CSD-Cascade manages interconnectedness through corrosive disadvantages and fertile functionings in an extended view of the capability approach. A -SD could either draw from its own ideology or other actualized -SD approaches; DSD may draw on Kittay's modified Rawlsian principles of justice as to whether a primary good is met, while drawing on notions of interconnectedness from CSD-Cascade through whether meeting a good or not has (dis)advantages. Finally, design requirements may be made through a value hierarchy and technical investigations (e.g., in-field prototype evaluations) should take place. The process, being iterative, should then restart at either the conceptual stage (if further information about the values are uncovered) or empirical stage (if more stakeholder involvement is needed). 

\section{Towards Other -Sensitive Designs}
As a meta-framework, different -SDs ought to iterate on one another; where one ethic may not include a concrete list of values or mechanisms for interconnectedness, other pre-existing concepts in -SDs (e.g., corrosive disadvantages and fertile functionings from CSD-Cascade) may be adapted or used in place. Rooted in Kittay's ethics, DSD builds off of this -SD meta-framework to create a design methodology for working with communities of people with disabilities or in dependant stages of life, that is participatory and value-based at every stage.  Through this discussion of DSD, we outline how a political philosophy or ideology may be adapted in a value-based approach to technology design. Moving beyond theories on social justice rooted in approaches such as Nussbaum and Kittay's, a -SD may go beyond the human as a primary stakeholder following philosophies similar to Bookchin's \emph{Social Ecology}~\cite{bookchinSocialEcologyCommunalism2007}. An \emph{Ecology}-Sensitive Design may adapt Bookchin's concepts of \emph{first} nature (ecological), \emph{second} nature (human social), and \emph{free} nature (harmonized) to technology design for technology for climate resilience, eco-activism, or social work.  We call for future work in value-based design to look at ideologies broadly and discover how they can be used to inform, guide, and act as a normative basis for incorporating the human (or beyond) in technology.


\bibliographystyle{ACM-Reference-Format}
\bibliography{sample-base}

\end{document}